# RECOMED: A Comprehensive Pharmaceutical Recommendation System

Mariam Zomorodi[1,*], Ismail Ghodsollahee[2], Jennifer H. Martin[3], Nicholas J. Talley[3], Vahid Salari[4], Pawel Plawiak[1,5], Kazem Rahimi[6], U. Rajendra Acharya[7]

[1] Department of Computer Science, Faculty of Computer Science and Telecommunications, Cracow University of Technology, Krakow, Poland

[2] Department of Computer Engineering, Ferdowsi University of Mashhad, Iran

[3] NHMRC Centre for Research Excellence in Digestive Health, Hunter Medical Research Institute (HMRI), The University of Newcastle, Callaghan, New South Wales, Australia

[4] Institute for Quantum Science and Technology, Department of Physics and Astronomy, University of Calgary, Alberta, Canada.

[5] Institute of Theoretical and Applied Informatics, Polish Academy of Sciences, Gliwice, Poland

[6] Deep Medicine, Nuffield Department of Women's and Reproductive Health, University of Oxford, Oxford, United Kingdom

[7] School of Mathematics, Physics and Computing, University of Southern Queensland, Toowoomba, QLD, Australia

**OBJECTIVES:** To build datasets containing useful information from drug databases and recommend a list of drugs to physicians and patients with high accuracy by considering features of a spectrum of people, diseases, and chemicals.

**METHODS:** A comprehensive pharmaceutical recommendation system was designed based on the features of a spectrum of people, diseases, and medicines extracted from two major drug databases, a dataset of patients and drug information. Then the recommendation was given, and the response was analysed using patient and caregiver ratings and the knowledge obtained from drug specifications and interactions. Sentiment analysis was employed by natural language processing approaches in pre-processing, along with neural network-based methods and recommender system algorithms for modelling the system. Patient conditions and medicine features were used for making two models based on matrix factorisation. Then we used drug interaction criteria to filter drugs with severe or mild interactions with other drugs.

We developed a deep learning model for recommending drugs by using data from 2304 patients as a training set, and then we used data from 660 patients as our validation set. We used knowledge from critical drug information and combined the model's outcome into a knowledge-based system with the rules obtained from constraints on taking medicine.

**RESULTS:** The results show that our recommendation system can recommend an acceptable combination of medicines according to the existing real-life prescriptions available. Compared with conventional matrix factorisation, our proposed model improves the accuracy, sensitivity, and hit rate by 26%, 34%, and 40%, respectively. In addition, it improves the accuracy, sensitivity, and hit rate by an average of 31%, 29%, and 28% compared to other machine learning methods. We have open-sourced our implementation in Python.

**CONCLUSION:** Compared to conventional machine learning approaches, we obtained average accuracy, sensitivity, and hit rate of 31%, 29%, and 28%, respectively. Compared to conventional matrix factorisation, our proposed method improved the accuracy, sensitivity, and hit rate by 26%, 34%, and 40%, respectively. However, it is acknowledged that this is not the same as clinical accuracy or sensitivity, and more accurate results can be obtained by gathering larger datasets.

## 1. INTRODUCTION

Recommendation systems (RS) are knowledge extraction systems that use information retrieval approaches to help people make better decisions and discover items through a complex information space [1], [2]. They have been around for many years, and with the advancement in machine learning approaches, their use has been widened, and it helps people to make more appropriate decisions in using different products. The popularity of using RS in different fields has increased since the announcement of the Netflix Prize competition that aimed to predict movie rates [3]. The recommender system application is extensive: from entertainment to e-commerce, the tourism industry, and medical recommender systems. Also, with rapid progress in artificial intelligence, there has been a greater acceleration in the application of recommender systems and their development.

The use of RS in medical science requires particular attention. The medical recommendation has some distinct features that make it special and must be used carefully. And because it affects people's health, there are many concerns about using RS for them. On the other hand, many people die every year because of medication errors. It has been reported as the world's third leading cause of death in [4], according to an Icelandic study. This makes the use of intelligent systems in medical science valuable and necessary. Drug prescription is also vital for physicians and involves considering different aspects, such as patient demographics, ethnicity, age, medication history, the specification of drugs for diseases related to the recommendation in question, and the medicine effectiveness for that specific case are among such concerns.

There are as many different medicines as 24000 [5] in just one database, however not all are for specific diseases or patients. Nevertheless a recommendation system to perform a set of suggestions for a particular patient with a specific disease might help physicians in considerations of prescribing the most appropriate medicines and also help patients to have a better choice in using drugs.

Recommender systems can be distinguished by the degree of risk imposed when a user accepts a recommendation [6]. In this regard, the medical domain can be seen as high risk, primarily due to the recommendation given to the user.
On the other hand, while having a comprehensive drug recommender system is important, designing an accurate system requires a dataset of drugs with patient outcomes, patient ratings, reviews, and drug information.

We gathered this information from two different and well-known databases Drugs.com [5] and Druglib.com [7], and we built three datasets which train the system and construct the final model. Finally, putting all of them together, we proposed a novel drug recommender system called RECOMED that learns the patient, medication history and drug features and also user reviews for different drugs to recommend a new drug to a patient.

The novelty of this work lies in the following parts:

1- Propose a pharmaceutical recommender system by considering the features of patients and medicines, including patients' conditions, age, gender, drug side effects, and drug categories.
2- Performing pre-processing steps on databases Druglib.com and Drugs.com websites to gather the appropriate data for our recommendation system, leading to comprehensive datasets for drug information.
3- Our system considers sentiment analysis of reviews, the prescriptions of doctors, and different similarity measures for recommending a medicine and its dose and other recommendations, including side effects and warnings for their usage.
4- Our system consists of a knowledge-based component to exclude drugs with serious side effects for a specific patient.
5- We proposed a model to predict the likely efficiency of medicines for patients.

The next section introduces the drug recommender systems and their challenges. Then in Section 3, we provide the current state-of-the-art recommender systems methods, specifically drug recommender systems. Section 4 elaborates on our proposed comprehensive drug recommender system in detail. Section 5 provides the results of this work, plus a discussion about them. Finally, in Section 6, we conclude the paper and present the future directions of this research.

## 2. RECOMMENDER SYSTEMS IN MEDICAL SCIENCE

One of the attractive and important applications of recommendation systems is medical recommendations and drug products. While they usually use the same recommendations

Here are the major differences between medical recommender systems and other recommender systems:

- Medical recommender systems care more about the health of patients than to make a profit.
- Security is the primary goal in drug recommender systems.
- Many existing recommender system techniques cannot be used, and others must be used cautiously because of safety issues.

- In the long term, time is considered an important factor in recommending a drug. In many situations, some drugs' negative effects are discovered over time. One example is zimeldine, with the increased risk of developing Guillain-Barre syndrome [18]. Rofecoxib, and Lumiracoxib are two more examples. Rofecoxib was withdrawn from the market about one year after marketing approval because of concerns about the increased risk of heart attack and stroke caused by using it in a long-term and high-dosage [19]. Lumiracoxib was also withdrawn due to the concern that it may cause liver failure [20]. So, a comprehensive medical recommender system should consider ratings in different time stamps to accurately estimate side effects discovered over time.

In the drug recommender system, the domain is medicine, and the exact contents to be recommended are one or more of the following lists:

1. A list of drugs > 1.
2. The dose - the amount of drug taken at one time.
3. The frequency at which the drug doses are taken over time.
4. Duration, which is how long the drug is taken.

Numbers 2 to 4 in the above list are referred to as the *dosage*. Therefore, we can define *a drug recommender system* as a smart system that is able to recommend a list of drugs plus their dosages with high accuracy in terms of a real prescription of a physician and also to have a positive effect on a patient, which available data can partially verify.

It should be noted that, of course, there is no medicine recommender system that we can trust thoroughly, and like other artificial intelligence systems applied in healthcare, their use and ethical issues must be addressed appropriately [21], [22].

## 3. LITERATURE REVIEW

Medical recommender systems have been around for many years, even before the emergence of recommender systems as a new field in computer science. According to [23], medicine recommender systems fall into two broad categories named "*ontology and rule-based approaches*" and "*data mining and machine learning-based*" approaches. Ontology-based recommender systems use the hierarchical organization of users and items to improve the recommendation [24].

Data mining and machine learning algorithms in the medical field are used to predict and recommend things like drug usefulness, having a disease [25], [26], the condition of the user, or ratings [27], [28]. For example, SVM, backpropagation neural network, and ID3 decision tree have been used in [29] for recommending drugs. The performance of these approaches has been compared in the above work, and the authors have shown that SVM has better accuracy and efficiency than the other algorithms. Their data set contains patients' age, sex, blood pressure, cholesterol, Na and K levels, and drug.

Some other researchers, while described as medical or medicine recommender systems, consider a detection and classification task where the trained dataset has some patient attributes, and based on that, the objective of the work is the detection or prediction of disease. Then for each disease, a set of medicines is recommended [29].

Sentiment analysis of drug reviews is one of the basic approaches for drug recommendations [30], [31], [32], [33], [34]. The sentiment analysis in these works mainly aims to recommend a drug or extract useful information like adverse drug reactions.

In [33], different deep learning approaches, such as CNN, LSTM, and BERT, have been investigated for sentiment analysis of patients' drug reviews. In another work, the combination of CNN-RNN has been applied.

In addition to recommendation systems, sentiment analysis and opinion mining of drug reviews is an active research area in drug review processing [35]. This analysis can be used for automatic opinion mining and recommending drugs.

A hybrid knowledge-based medical prescription approach has been presented in [36]. The authors use historical medical prescriptions to recommend a list of medicines to physicians. The approach uses the similarity between cases where a case is a medical information like demography, treatment, age, sex, symptoms, and diagnosis. Based on the degree of similarity, a drug list is produced. The list is complemented by Bayesian reasoning, where a model of the conditional probability of drugs is built. This approach has been applied in Humphrey & Partners Medical Services Limited medical centre.

Some works in medical recommendation have focused on particular drugs like diabetes [37]. Their model is based on the ontology of medical knowledge and a decision decision-making approach for multiple criteria and computes the medication. Then by using the entropy, the information about patient' history has been computed, and finally, the most appropriate medications have been recommended to the physicians.

Many recommender system approaches have not been well considered in the medical and pharmaceutical recommendations. However, using polarity in sentiment analysis of user comments is one of the important parts of using NLP in recommendation systems. It can be viewed as determining whether a word or phrase in the document or even a whole document is positive, negative, or neutral in general.

Figure 1 shows the broad classification of different recommendation system approaches in pharmaceutical research. We can see a growing tendency to use machine learning approaches in this field.

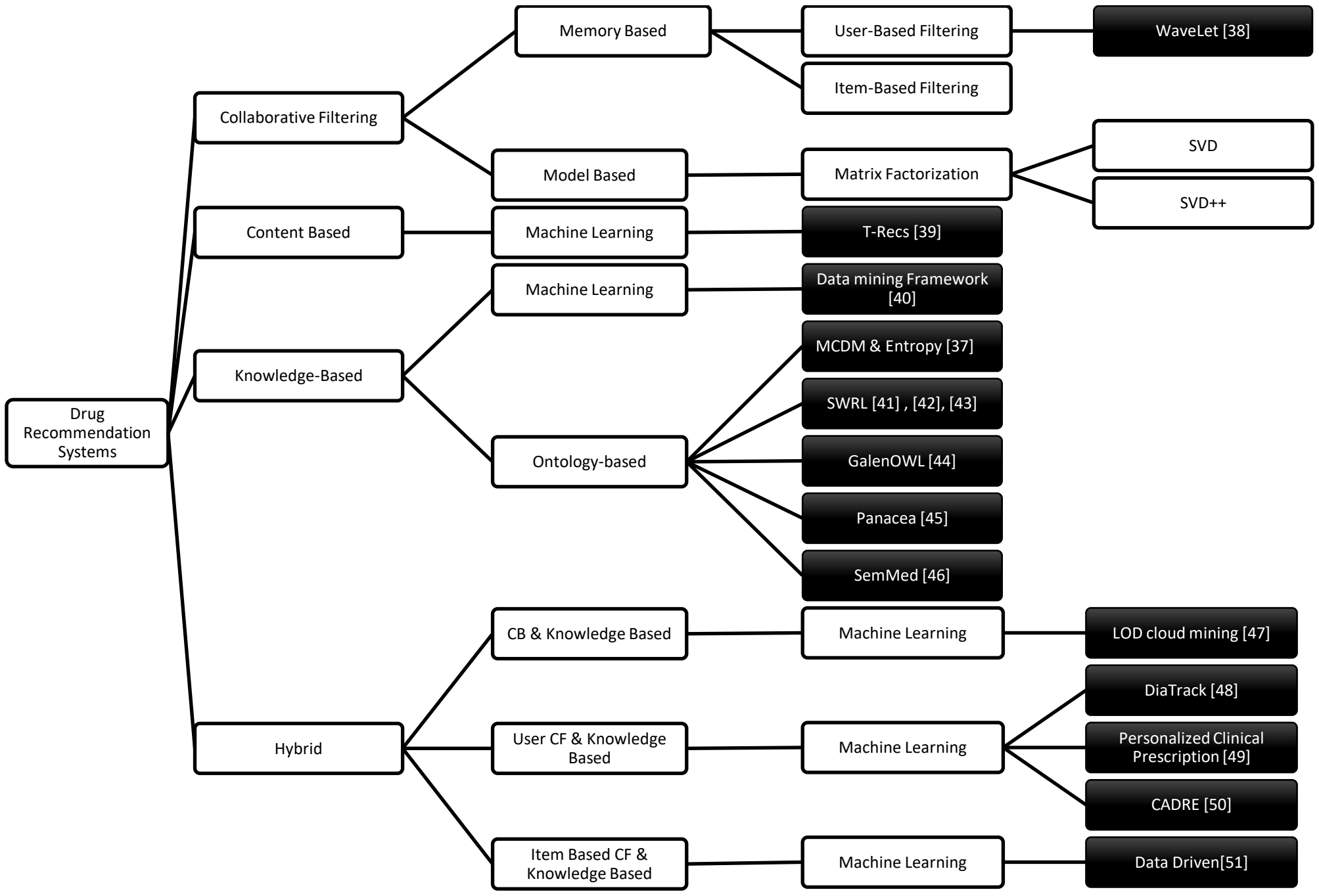

**Figure 1: Broad classification of recommendation system techniques.**

## 4. Material and methods

In this section, we formulate the medicine recommender system problem and present our approach for the general medicine recommender system.

Many recommendation systems, like collaborative filtering and content-based approaches, mainly rely on past information to make decisions for the current situation. It is not always the case in the domain of drug recommendation. The patient’s condition is different compared to the other patients and compared to the same patient over time. So, in addition to the history information like general rates, reviews, and the effective rate of the drug, it is necessary to use the patient's current condition to make a more accurate decision. We also cannot rely on diversity-based recommendations as it is used in some recommender systems, like the one used in Netflix; even if the drug is not rated high, it can be suitable for some patients.

On the other hand, many recommender systems rely on user knowledge; when there is a lack of user knowledge, we cannot personalise them. While we have an adequate dataset for our recommendation task, the problem emerges when new inputs enter the system. In our medicine recommender system, these inputs can be new patients or new drugs. *Cold start* problem is a term used for this problem, and it is a challenging issue in designing any recommender system. We reduced this effect by applying a clustering-based approach. Because drugs are clustered into a specific category, we can put a new drug in the category which belongs to it, so we use the same rating for the new drugs as those in that category. This is effective in solving the cold start problem in our recommender system.

**Proposed method**

This section discusses all phases of our model for building RECOMED system.

We present a novel hybrid drug recommender system (RS) with features of several recommender systems. The system uses natural language processing (NLP) and other machine learning techniques to implement the system. The proposed approach uses CB, CF model-based, knowledge-based, and deep learning for drug recommendation. Here in this section, we elaborate on each step toward the final drug recommendation for each patient. After a very intensive web crawling through two well-known pharmaceutical websites, Drugs.com and Druglib.com, and building three different datasets, feature extraction and modelling are performed. Then in the next step, recommendations for proper drugs are provided. Detailed implementation of this stage is given in section 5.4. At the final stage, the list of drugs is refined based on defined rules and the ratings and drug features which is an important aspect of our medicine recommender system.

Figure 2 represents the whole RECOMED model in the training stage of our work consists of four components, and we elaborate on each phase of our approach in more detail in the following parts.

The first two components deal with dataset extraction and preprocessing steps and the second two deal with modelling the recommendations and using the final knowledge-based decision. Figure 3 is a graphical representation of the first two parts.

### 4.1. Dataset extraction

In this work, any recommendation for drugs and their dosage is based on the patients' features like age, gender, previous illness, and other drugs they consume, and drugs' features like drug classification, side effects, and drug interactions. So, in this phase, the extraction of user features, drug features, and drug interaction datasets from Drugs.com and Druglib.com databases is accomplished. In the second step of this phase, the dataset is prepared for clustering and modelling the recommendation system. The review field in the drug recommendation database contains users' and caregivers' opinions about drugs' effectiveness. Based on our information, none of the existing datasets have complete and comprehensive patient and drug information.

We built three different datasets named *users*, *drugs*, and *interactions*.

### 4.2. Drugs and users datasets

Druglib.com and Drugs.com were employed in this work to extract information about patients and drugs and build two datasets named drugs and patients. We should mention that there are also other databases for drug information and recommendations, like SIDER [52], for drug side effects. We intend to include them in future works to build a complete drug dataset. Three features side effects, benefits, and membership in a given drug category were considered for drugs.

First, different drug categories and side effects were extracted in tables 1 and 2. There are 150 different drug categories, and 128 side effects were extracted from the Druglib.com database.

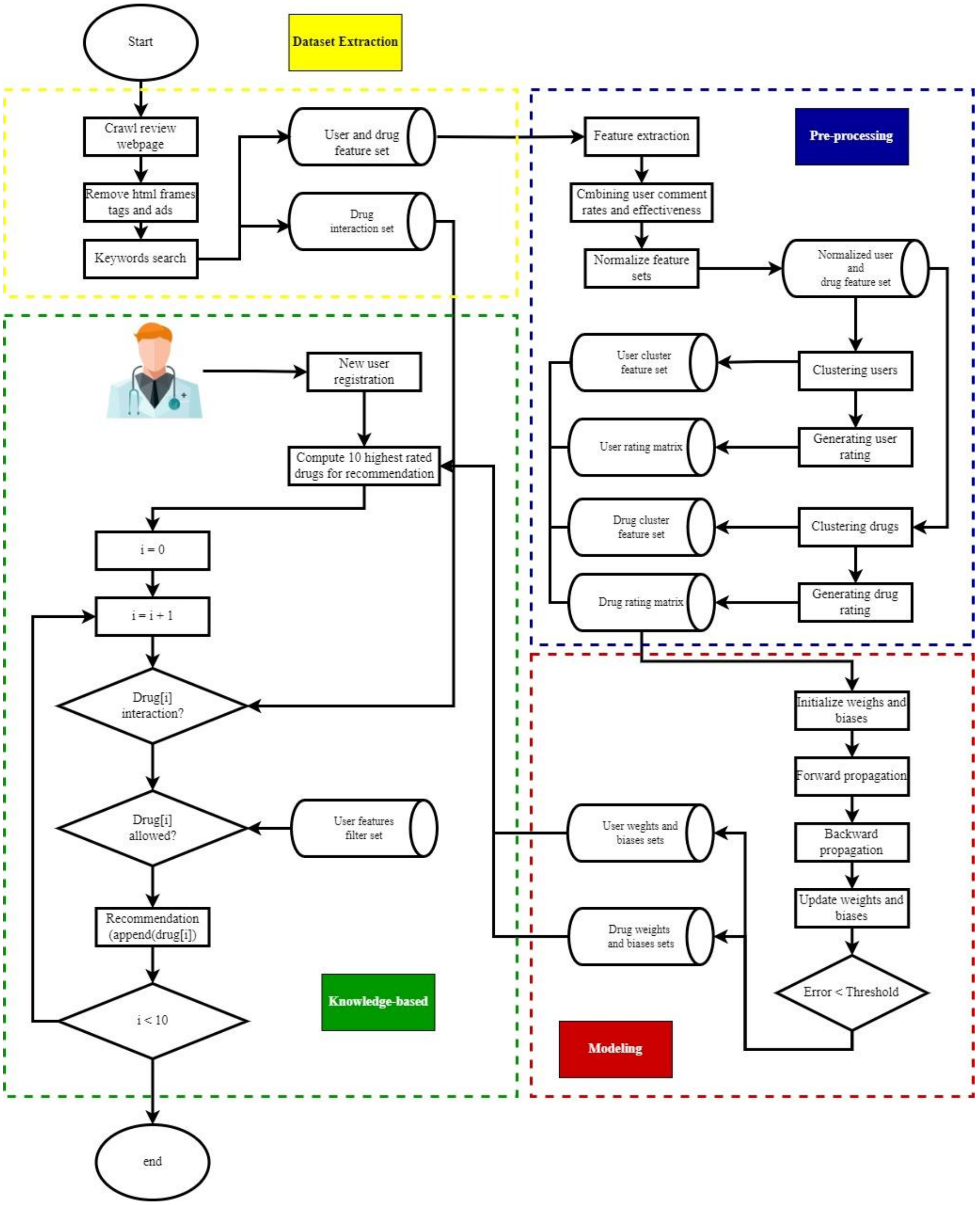

**Figure 2: Components of RECOMED drug recommendation system in the training stage.**

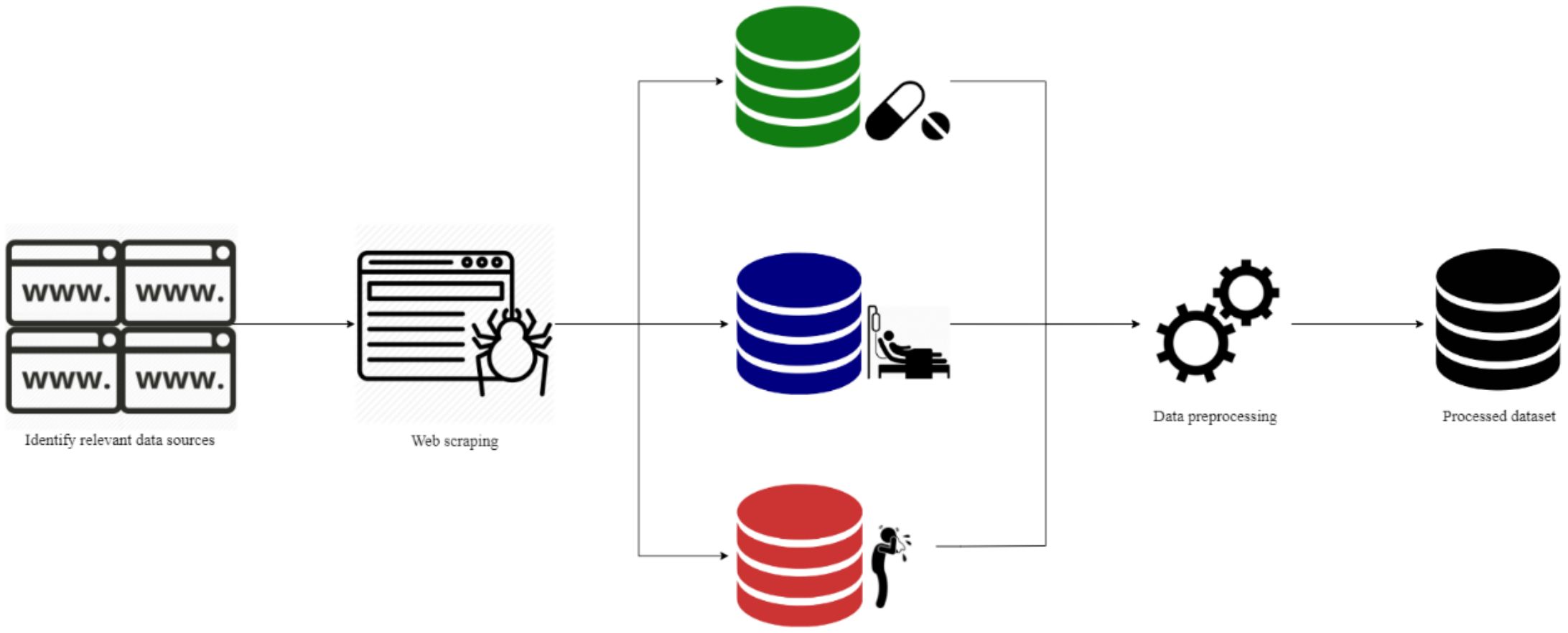

**Figure 3- Dataset extraction and preprocessing in our proposed approach**

**TABLE1 - DRUG CATEGORIES LIST**

| Index | Category |
|---|---|
| 1 | Acetylcholine-Agonists |
| 2 | Adrenergic-Alpha-Agonists |
| … | |
| 150 | Vasodilators |

**TABLE2 -DRUGS SIDE EFFECTS**

| Index | Side Effects |
|---|---|
| 1 | Completed suicide |
| 2 | Confusional state |
| … | |
| 128 | Wrong drug administered |

Then drug *benefits* were also extracted and combined with the information in the above tables, and finally, the *drugs* dataset was prepared, as is partially shown in Table 3.

**TABLE 3- DRUGS DATASET**

| Index | Drug Name | Drug Category | | | | Side Effects | | | | Benefits | | | |
|---|---|---|---|---|---|---|---|---|---|---|---|---|---|
| | | 1 | 2 | … | 150 | 1 | 2 | … | 128 | 1 | 2 | … | 88 |
| 1 | Hytrin Terazosin | 1 | 0 | … | 0 | 0 | 0 | … | 0 | 1 | 1 | … | 0 |
| 2 | Mirtazapine | 1 | 0 | … | 0 | 1 | 1 | … | 0 | 0 | 0 | … | 0 |
| … | … | … | … | … | … | … | … | … | … | … | … | … | … |
| 480 | Proscar Finasteride | 0 | 0 | … | 1 | 0 | 0 | … | 0 | 0 | 0 | … | 0 |

We also extracted the *users* dataset of patient features and comments on different drugs. Six features are considered for *users* datasets: age, gender, current disease (condition), other conditions, other drugs are taken, and user level, which is patient or caregiver.

Table 4 represents the structure of this dataset.

**TABLE3- USERS DATASET**

| index | Level | Age | Genus | Condition | Other Condition | Other Drug | Drug Name | Overall Rating | Effectiveness | Side Effects | Comment |
|---|---|---|---|---|---|---|---|---|---|---|---|
| 1 | Patient | 22 | Male | Depression | Sleeplessness | None | Mirtazapine | 1 | Ineffective | Severe Side Effects | … |
| 2 | Patient | 38 | Male | Depression | None | None | Mirtazapine | 2 | Ineffective | Moderate Side Effects | … |
| … | ... | ... | ... | ... | ... | ... | ... | ... | ... | ... | ... |
| 3294 | Patient | 28 | Male | Hair loss | None | None | Proscar Finasteride | 4 | Moderately Effective | Mild Side Effect | … |

### 4.3. Interactions dataset

The last dataset prepared in this work is the *interactions* dataset. This information is important for recommending the appropriate medicine list to the patients. We extracted drug interaction information from Drugs.com, and after mapping drugs' names with their counterparts in Druglib.com, the interaction dataset, partially presented in Table 5, was created with 180 drug interaction information.

**TABLE4 -DRUG INTERACTION DATASET**

| | Abilifish | … | Cimbaita | … | Syntroid | … | zyban |
|---|---|---|---|---|---|---|---|
| Abilifish | … | … | Moderate | … | .. | … | … |
| Accupril | … | … | … | … | Moderate | … | … |
| Aciphex | … | … | Major | … | - | … | … |
| … | … | … | … | … | … | … | … |
| Zyban | … | … | … | … | … | … | … |

### 4.4. Dataset preparation

In this phase, our dataset is prepared for creating the recommendation model in the next step. First, using Natural Language Processing (NLP) techniques, user and drug features are extracted, and then normalisation and clustering are accomplished to prepare the datasets for modelling the recommendation system. Here, we elaborate on each of these steps:

### 4.5. Feature extraction

The first pre-processing step is feature extraction from user feature and drug features datasets. Bag-Of-Words (BOW) method is used for this purpose.

**NLP for extracting drug and user features**

The feature extraction was mainly performed using natural language processing (NLP) techniques. Two well-known methods to extract text features by NLP are Bag-of-Words (BOW) and term frequency-inverse term frequency (TF-IDF).

Our proposed pharmaceutical recommendation system uses the BOW feature extraction method to perform feature extraction from database texts. This method consists of four steps:

- Text pre-processing
- Vocabulary creation
- Building feature matrix
- Polarity of user comments

Here, every part of this process has been described:

### 4.6. Text pre-processing

In the text pre-processing step, all punctuations and symbols are removed, and abbreviations are converted into their full names or phrases. Some of these conversions are presented in Table 6. Moreover, spelling mistakes were corrected using the TexBlob library of Python, and stop words were removed using a predefined list of stop words.

**TABLE 6- EXAMPLES OF ABBREVIATIONS TO FULL NAME CONVERSIONS**

| Abbreviations | Original Form |
|---|---|
| HBP | high blood pressure |
| COPD | chronic obstructive pulmonary disease |
| PMS | premenstrual syndrome |
| OCD | obsessive-compulsive disorder |

## Vocabulary creation

Using NLP techniques, a vocabulary of words is created in the second step of feature extraction. For this purpose, an array of words is created by checking all registered words in the dataset. This array is constructed from unique words of the dataset and their frequency. Words are rearranged according to their frequency to deal with the random filling of the feature matrix. Moreover, to deal with the sparseness of the feature matrix, words with low frequency are removed. Some of the most frequent words extracted from the datasets created and discussed in the previous section can be seen in Table 7.

**TABLE 7. EXAMPLES OF MOST FREQUENT WORDS IN DATASETS.**

| Term | Frequency |
|---|---|
| Pain | 33 |
| Infection | 22 |
| Surgery | 15 |
| Chronic | 13 |

## Building feature matrix

The feature matrix is created in the third step of extracting the features. For this purpose, a unique word is assigned to each column of the matrix, and a new row is considered for each user review. Each cell of this matrix represents the existence of the word in the user’s review, which is essentially zero or one.

## Polarity of user comments (PUC)

We used NLP and opinion mining to extract PUC. This approach aims at extracting the opinion of users as a positive or negative comment. The output of this component is used in the users’ rating matrix.

**ALGORITHM1. COMMENT POLARITY ACQUISITION**

| |
|---|
| Input: $UserComments, StopWords$ |
| Output: $PolarityOfUserComments$ |
| 1. $Removing\ Capital\ Letters\ and\ Emojis$ from $UserComments$ |
| 2. $Removing\ StopWords\ from\ UserComments$ |
| 3. $Word\ Tokenize\ (UserComments)$ |
| 4. $Word\ Lemmatization(UserComments)$ |
| 5. $Frequency\ of\ Words(UserComments)$ |
| 6. $TextBlob(UserComments)$ |

## Combined User Rating Acquisition

To have a more accurate rating for drugs, we considered the combined user comments and ratings from different sources. This overall rating is called Combined User Rating Acquisition (CUR) parameter and is obtained from analysing user comments and ratings as follows:

1. $Overall\ Rating\ \in Z, 0 \leq Overall\ Rating\ \leq 10$
2. $Effectivness\ \in E, E = \{$Ineffective, Marginally Effective, Moderately Effective, Considerably Effective, Highly Effective$\}$
3. $Side\ Effect\ \in S, S = \{No\ Side\ Effect,\ Mild\ Side\ Effect, Moderate\ Side\ Effect,\ Severe\ Side\ Effect,\ Extereml\ Severe\ Side\ Effect\}$
4. *User Comment*

CUR parameter is calculated as equation (6), and the above parameters are replaced by CUR in the user feature matrix:

$$\mathrm{CUR} = \frac{\frac{\left(\frac{OverallRating}{10}+\frac{DOE}{4}\right)}{2} - \frac{DOS}{4} + PUC}{2} \qquad (6)$$

In equation (6), *DOE (Degree of Effectiveness)* represents the degree of drug effectiveness. The user selects the effectiveness of a drug from a list of five different options: *Ineffective*, *Marginally Effective*, *Moderately Effective*, *Considerably Effective*, and *Highly Effective*, and it takes a number in the range [0-4]. Similarly, DOS (Degree of Side Effects) is the degree that a drug has a side effect (range [0-4]), and the numbers applied in the denominator are for normalization purpose.

PUC (Polarity of User Comments) is calculated using Natural Language Processing (NLP), and opinion mining techniques and the nltk library in Python are used in this regard. Algorithm 1 shows the steps of the work for calculating PUC.

**Normalization-** After extracting features from drug and user datasets, these features should also be normalized to perform better in training the model.

| Index | Level | Age | Genus | Condition | Other Condition | Other Drug | Drug Name | Overall Rating | Effectiveness (DOE) | Side Effects (DOS) | Comment |
|---|---|---|---|---|---|---|---|---|---|---|---|
| 1 | patient | 22 | male | depression | Sleeplesness | None | Mirtazapine | 0.1 | 0 | 0.75 | 0.8 (Obtained from PUC) |
| … | … | … | … | … | … | … | … | … | … | … | … |
| 3294 | patient | 28 | male | Hair loss | None | None | Proscar | 0.4 | 0.5 | 0.25 | 0.2 |

| index | Level | Age | Genus | Condition | Other Condition | Other Drug | Drug Name | Combined User Rating (CUR) |
|---|---|---|---|---|---|---|---|---|
| 1 | patient | 22 | male | depression | Sleeplesness | None | Mirtazapine | 0.05 |
| … | … | … | … | … | … | … | … | … |
| 3294 | patient | 28 | male | Hair loss | None | None | Proscar | 0.1 |

**Figure 4- Combination of different user ratings for a given drug**

Figure 4 is the final user rating dataset after applying the combined user rating acquisition stage. This stage converts the dataset on the left side into the right-side dataset. Each column in both datasets has a given user's features along with the drug name they rate. In the left-side dataset, we can see different user ratings, and then in the right-side dataset, these ratings are combined into CUR using equation (6).

### 4.7. Clustering

Clustering is considered one of the main steps in a recommender system for improving diversity, consistency, and reliability [53]. Clustering has been considered in many works in recommender systems, particularly for reducing the sparsity of data [54], [55]. Due to the sparseness of the rating matrix, we consider a clustering-based approach, and patients are clustered before performing the matrix factorisation, which is elaborated in the next part. This clustering is mostly required because users usually review only one drug corresponding to a specific disease, so the rating matrix is highly sparse. Clustering can help group the users and drugs with similar features and significantly resolve the sparsity problem. Users are clustered based on their gender, age, comments, and being patient or caregiver. It is clear that after clustering, each class of users reviews several drugs, which can improve the matrix factorisation process. In the same way, drugs are also clustered based on their side effect, effectiveness, interaction, etc.

We used a modified K-means algorithm [56] to perform this clustering. While the original K-means algorithm is unsupervised, used for clustering, the number of clusters is pre-determined. So it couldn't be utilized in the same way in our proposed drug recommendation system. Therefore, in this paper, we employed the U-Kmeans method [56]. This method performs the unsupervised K-means and determines the best cluster numbers that lead to better classification performance. Each row of the dataset and the centre of each cluster are represented by F= $\{f_1, \dots, f_n\}$ and A= $\{a_1, \dots, a_k\}$ respectively and the K-means objective function is defined as (7).

$$J(M, A) = \sum_{i=1}^{n} \sum_{j=1}^{k} M_{ij} \left\| f_i - a_j \right\| \qquad (7)$$

Where in (7), $k$ is the number of clusters, $n$ is the number of dataset features, and $M_{ij}$ indicates the membership of $F_i$ to the $j_{th}$ cluster. In the K-means algorithm, this objective function must be minimized. In [57], an entropy-based method is proposed to improve K-means. In this method, to determine the centers of the clusters, Equation (8) is added to the objective function.

$$B_n \sum_{j=1}^{k} a_j \ln a_j \qquad (8)$$

In (8), the effect of the cluster imbalance is added to the objective function. As can be seen in (9), when the $B_n$ coefficient of the improved objective function is zero. The following K-means objective function is obtained.

$$J(M, A) = \sum_{i=1}^{n} \sum_{j=1}^{k} M_{ij} \left\| x_i - a_j \right\| - B \sum_{j=1}^{k} \eta_j \ln \eta_j \qquad (9)$$

Where in this equation, $\eta_j$ represents the number of members of a cluster, which is determined by (10).

$$\eta_j = \frac{\sum_{i=1}^{n} M_{ij} x_i}{\sum_{i=1}^{n} M_{ij}} \qquad (10)$$

In [56], equation (11) is considered to determine the optimized number of clusters. By adding this term to equation (10), the final objective function is obtained as (12).

$$\mathrm{L} \sum_{i=1}^{n} \sum_{j=1}^{k} M_{ij} \ln a_j \qquad (11)$$

$$J(M, A, a) = \sum_{i=1}^{n} \sum_{j=1}^{k} M_{ij} \left\| x_i - a_j \right\| - B \sum_{j=1}^{k} a_j \ln a_j - \mathrm{L} \sum_{i=1}^{n} \sum_{j=1}^{k} M_{ij} \ln a_j \qquad (12)$$

The pseudocode of the U-K-means classification method based on the approach in [56] is presented in Algorithm 2.

**Algorithm2 . Our modified Pseudo code of U-Kmeans based on [54].**

$initial\ c^{(0)} = n, \alpha_k^{(0)} = \frac{1}{n}, a_k^{(0)} = x_i$

$Initial\ learning\ rates\ L^{(0)} = B^{(0)} = 1$

$Set\ t = 0\ , \varepsilon > 0$

$while\ max \| a_k^{t+1} - a_k^t \| < \varepsilon$

If $\| x_i - \alpha_k \|^2 - Lln\alpha_k = \min_{1 \le k \le c} \| x_i - a_k \|^2 - Lln\alpha_k$

$M_{ik}^{(t+1)} = 1$

Else

$M_{ik}^{(t+1)} = 0$

$L^{(t+1)} = e^{-c(t+1)/250}$

$\alpha_k^{(t+1)} = \sum_{i=1}^{n} \frac{M_{ik}}{n} + \left(\frac{B}{L}\right) \alpha_k^{(t)} \ln a_k^t - \sum_{s=1}^{c} \alpha_s^t \ln a_s^t$

$B^{t+1} = min\left( \frac{\sum_{k=1}^{c} \exp\left(-\eta^n \left| a_k^{t+1} - a_k^t \right|\right)}{c}, \frac{1 - \max_{1 \le k \le c}\left(\frac{1}{n} \sum_{i=1}^{n} M_{ik}\right)}{-\max_{1 \le k \le c} a_k^t \sum_{k=1}^{c} \ln a_k^t} \right)$

$update\ C^t\ to\ C^{t+1}\ by\ discard\ those\ cluster\ with\ a_k^{t+1} \le \frac{1}{n}$

$a_k^* = \frac{a_k^*}{\sum_{s=1}^{c(t+1)} a_s^*}$

$M_{ik}^* = \frac{M_{ik}^*}{\sum_{s=1}^{c(t+1)} M_{is}^*}$

$a_k = \frac{\sum_{i=1}^{n} M_{ik} x_{ij}}{\sum_{i=1}^{n} M_{ik}}$

$if\ t \ge 60\ and\ c^{(t-60)} - c^t = 0$

$B^{(t+1)} = 0$

t=t+1

**TABLE 8: RATE MATRIX WITHOUT CLASSIFICATION**

| Users \ Drugs | 1 | 2 | 3 | 4 | 5 | 6 | … | 615 | 616 | 617 | 618 | 619 |
|---|---|---|---|---|---|---|---|---|---|---|---|---|
| 1 | 1 | | | | | | | | | | | |
| 2 | | | | | | | | 3 | | | | 3 |
| 3 | | | | | | | | | 5 | | | |
| … | | | | | | | | | | | | |
| 979 | | 1 | | | | | | | | | | |
| 980 | | | | | | | | | | | 3 | |
| 981 | | | | | | 3 | | | | | | |

**TABLE 9: RATE MATRIX AFTER CLASSIFICATION**

| Users \ Drugs | 1 | 2 | 3 | 4 | 5 | 6 | … | 615 | 616 | 617 | 618 | 619 |
|---|---|---|---|---|---|---|---|---|---|---|---|---|
| 1 | 1 | | | | | | | | | | | |
| 2 | | | | | | | | 3 | | | | 3 |
| 3 | | | | | | | | | 5 | | | |
| … | | | | | | | | | | | | |
| 38 | | 1 | | | | | | | | | | |
| 39 | | | | | | | | | | | 3 | |
| 40 | | | | | | 3 | | | | | | |

## 4.8. Modelling

In the next step, the clustering outcome is used to build a recommender system model able to recommend the best drugs. Later, we filter the model's output with a knowledge-based component for safety reasons.

### Neural Network-based Matrix Factorization

Matrix factorization is a popular method for recommender systems aiming at finding two rectangular matrices called user and item matrices with smaller sizes than the rating matrix [58]. The dot product between these two matrices results in the rating matrix.

To reduce the computational overhead, cope with the sparsity of the ratings, and increase accuracy, we proposed a neural network-based matrix factorization technique. The first two matrices, Rating and Effectiveness, are constructed by extracting information from Druglib.com.

In our model, the rating matrix $Rating \in \mathrm{R}^{n*m}$ is estimated as the multiplication of two matrices $Clusters^{n*k}$ and $Drugs^{k*m}$ as (13):

$$Raing \approx \mathrm{Clusters}.Drugs^{T} \quad (13)$$

This model applies a neural network algorithm to estimate the users' comments for each medicine. Clustered users and drugs and also users' and drugs' features are used in building this new model, as illustrated in Figure 5.

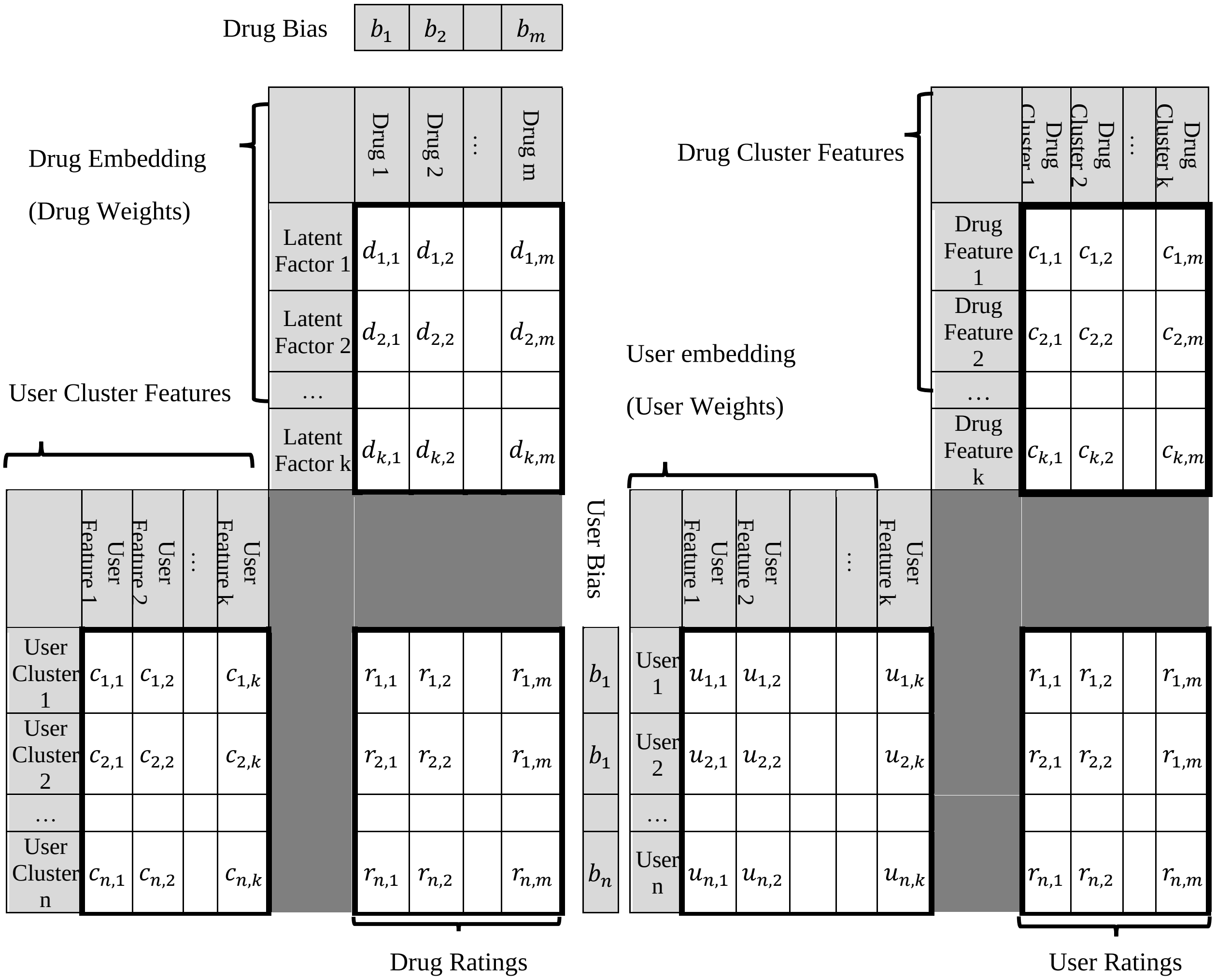

**Figure 5- Our proposed customized matrix factorization method.**

The input to the neural network is user and drug-clustered features. Drug Embedding and User Embedding matrices are the input to this network. The output of the network is two ratings. One of them resulted from the neural network MF with classified user features, and the other one is from the neural network MF with classified drug features. With sparse rating matrices, the forward and backward pass calculations are accomplished just for non-zero ratings to reduce the computation load. The neural network layer output is calculated as:

$$a^{z+1} = f^{z+1}\left(\sum_{i=1}^{K} w_i^{z+1}.\psi^{z+1}(n,m).a_i^z + b_j^{z+1}\right) \quad i \in (1,K), j \in (1,MN), z \in (0,Z-1), n \in (0,N), m \in (0,M) \qquad (14)$$

In this equation, $f^{z+1}$ is the activation function, $w_i^{z+1}$ are the weights, $b_j^{z+1}$ are the biases, $Z$ is the number of layers, $M$ is the number of drugs, $N$ is the number of users, $a^Z$ is the output of the network, $K$ is the number of the features for drugs or users, and $MN$ represents the number of drugs in the user network and represents the number of users in the drugs network. And finally $\psi^{z+1}$ is the rating existence function defined as:

$$\begin{cases} \psi^{z+1}(n,m) = 1 & if(Rating\ n,m\ \ exit\ or\ z < Z-1) \\ \psi^{z+1}(n,m) = 0 & if(Rating\ n,m\ not\ exist) \end{cases} \qquad (15)$$

Also, the backward pass calculations are as equations (16) to (19) for the output and hidden layers respectively:

For the output:

$$\Delta out = \left(R_{(n,m)} - a^Z\right).\psi^Z(n,m).f^{z+1'}(a^Z) \quad n \in (0,N), m \in (0,M) \qquad (16)$$

$$\Delta W^Z = \Delta out.a^Z.\gamma^Z \qquad (17)$$

For the hidden layers:

$$\Delta Hidden^z = f^{'}(a^z).\sum_i \Delta out_i\,.w_i^Z \qquad i \in (1,K), z \in (0,Z-1) \qquad (18)$$

$$\Delta w^z = \Delta Hidden^z.a^z.\gamma^z \qquad z \in (0,Z-1) \qquad (19)$$

In these equations, $f^{z+1'}$ is the gradient of the activation function, $R_{(n,m)}$ is the rating corresponding to the users or drugs, $\Delta W^Z$ is the error correction for the output layer, $\Delta w^z$ are the error corrections for the hidden layers, and $\gamma^z$ is the learning rate. The weight updates are also according to equation (20):

$$w_{new}^{z} = w_{old}^{z} + \Delta w^{z} \; z \in (0, Z) \quad (20)$$

### 4.9. Knowledge-based component

After modeling the recommendation system, several constraints on the model output are applied. The final stage in the recommendation process is based on the knowledge-based technique. The knowledge-based recommendation is a specific recommender system that can be used in combination with other algorithms or alone.

The aim of using this module is its huge impact in increasing the safety of the recommendations.

We extracted and gathered rules in the drug recommendation domain as queries. These rules are based on *Drug Interactions* and *Adverse Events*. Using these rules, we can prevent recommending drugs that lead to events like death, hospitalization, disability, and life-threatening events. The flowchart of this component has been extracted from Figure 2 and redrawn in Figure 6.

The set of these rules which our knowledge-based component considers falls into these two categories:

- Based on patients' features:
    - Gender is allowed to recommend a drug.
    - The age is allowed for recommending a drug.
- Based on drug interactions:
    - The recommended drug has no interaction with other drugs taken by the user.

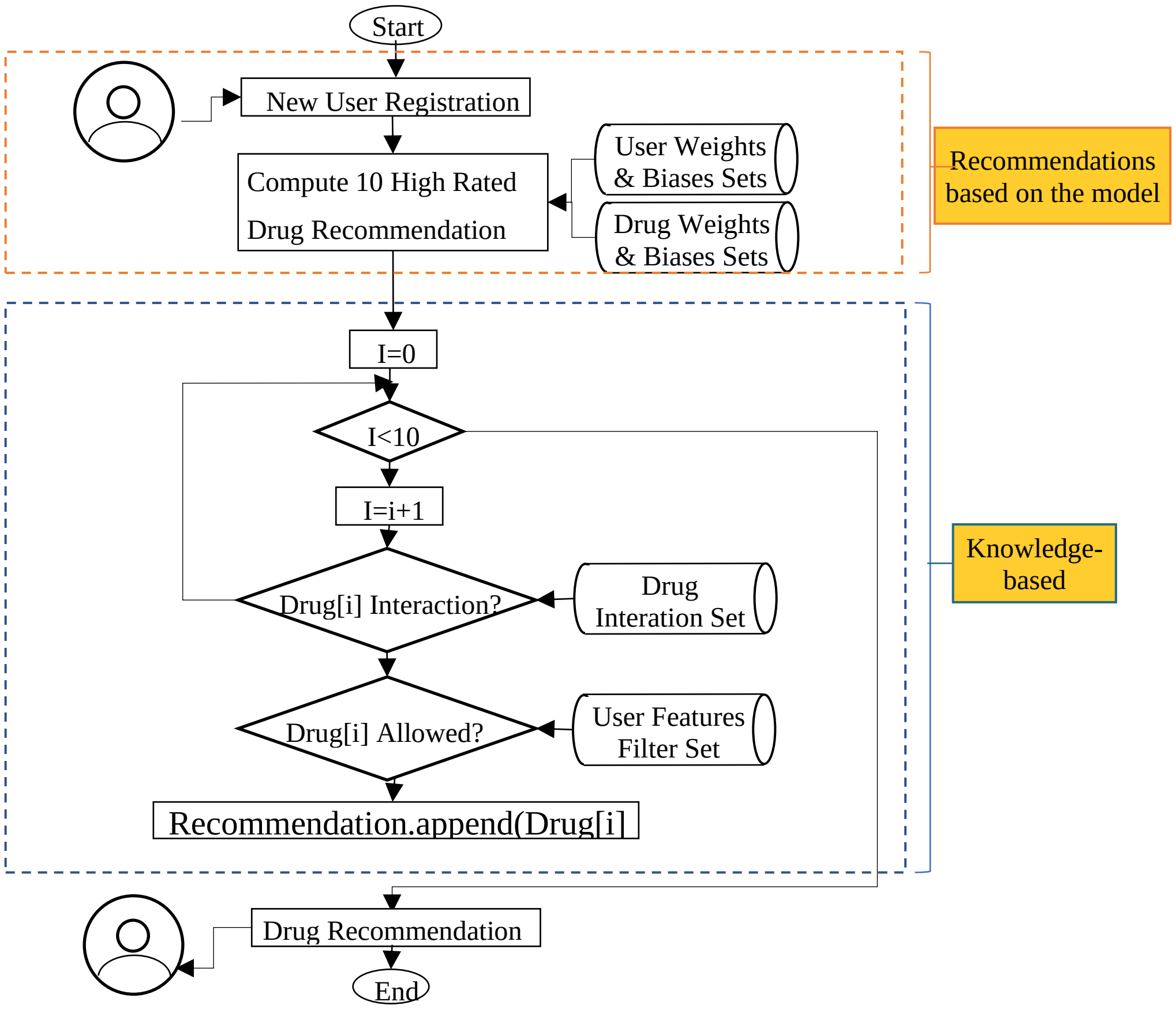

**Figure 6- Knowledge-based component of our proposed approach based on the bottom left of Figure 2.**

Table 10 presents knowledge-based rules based on patient's features that have been considered in this work. For example, according to this table, a drug can only be recommended if the patient's age is in the allowed range and the gender is allowed for recommending the drug.

**TABLE 10- USER FEATURE-BASED RULES**

| Drug Name | | Abilify | Actemra | | Zometa |
|---|---|---|---|---|---|
| Not allowed age ranges | Minimum | 24 | 29 | … | 31 |
| | Maximum | 45 | 64 | … | 67 |
| Allowed gender | None | 0 | 1 | … | 0 |
| | Female | 0 | 0 | … | 0 |
| | Male | 0 | 0 | … | 0 |
| | Both | 1 | 0 | … | 1 |

For our proposed knowledge-based component, another adverse events dataset is generated from Druglib.com. The structure of this dataset is presented in Table 11. Features in this dataset include age, gender, the name of the drug taken by a given patient, its adverse event, reaction, and other drugs used by the patient.

**TABLE11-ADVERSE EVENTS DATASET**

| Index | Drug Name | Age | Genus | Reaction | Adverse Event | Other Drug |
|---|---|---|---|---|---|---|
| 1 | Ability (Airipiprazole) | 63 | male | | Death | - |
| … | ... | … | … | … | … | ... |
| 12 | Acterma | 47 | female | | Death; Hospitalization | Insulin |
| … | … | … | … | ... | … | … |
| 2486 | Zyprexa | 50 | male | | Hospitalization | Fluticasone propiate; Salmeterol; Carbemazepine |

We used Gaussian and Poisson distribution for patients' age and gender from the above dataset for the adverse events of using a specific drug. These adverse events can be death, hospitalization, disability, or other life-threatening events.

Since, in this case, we require the average and standard deviation, by using Poisson and Gaussian distribution, it is possible to compute the allowed gender for recommending a drug to a patient using much less memory than machine learning for this specific task.

Assume that on average, by recommending a drug $\gamma$ for $\eta$ times to patients with $gender = female$ they experience one of the adverse events mentioned in Table 11, then the probability that by recommending drug $\gamma$ to a female patient she experiences one of the adverse events is calculated as equation (21):

$$P_n(x) = e^{-\lambda_{Female}^{\Upsilon}} \lambda_{Female}^{\Upsilon} \quad \lambda_{Female}^{\Upsilon} = \frac{Number\ of\ Adverse\ Event}{\eta} \qquad (21)$$

And similarly, for a male patient, this probability is calculated as equation (22):

$$P_n(x) = e^{-\lambda_{Male}^{\Upsilon}} \lambda_{Male}^{\Upsilon} \quad \lambda_{Male}^{\Upsilon} = \frac{Number\ of\ Adverse\ Event}{\eta} \qquad (22)$$

Using the above calculations, if the probability of an adverse event for each gender and each medicine is more than a given threshold value, the medicine is removed from the list and is not recommended to the patient. In this paper, we set the $threshold = 50\%$.

Also, normal distribution was used for setting the rules related to the patients ages. Suppose the average and standard deviation of a patient's age who have taken medicine $\gamma$ and has an adverse event is represented by $\mu$ and $\sigma$, respectively. In that case, the normal distribution function related to age is as equation (23):

$$f(x) = \frac{1}{\sqrt{2\pi}\sigma_{\Upsilon}} e^{-\frac{1}{2}\left(\frac{x-\mu_{\Upsilon}}{\sigma_{\Upsilon}}\right)^2} \qquad (23)$$

and so for patients who are taking medicine $\gamma$, equation (17) for age range has to be met to minimize the adverse event (24):

$$X \epsilon \left(\mu_{\Upsilon} - 1.96\left(\frac{\sigma_{\Upsilon}}{\sqrt{n}}\right), \mu_{\Upsilon} + 1.96\left(\frac{\sigma_{\Upsilon}}{\sqrt{n}}\right)\right), 1 - a = 95\%, Z_{0.975} = 1.96 \qquad (24)$$

In this research, we used the rules related to the users' features and the medicine rules and drug interactions we are also considering. In this regard, the drug interactions dataset was used to exclude recommendations for drugs having high interactions with other drugs.

## 5. RESULTS

This section discusses our proposed drug recommendation system implementation and the newly generated datasets. First, we explain the extracted and newly generated datasets and then we will demonstrate the results of our implemented system.

### The dataset

As discussed in the proposed method, we used the information from two databases of drugs Druglib.com [7] and Drugs.com [5]. The first database Druglib.com is a comprehensive resource for drug information. For each drug, a variety of information such as description, side-effects, drug ratings & reviews by patients, and clinical pharmacology has been provided. Also, Drugs.com is another database for drug information, and many recommendation systems have been suggested that use this database to build their models. Both the original and the revised version of Drugs.com have been used in RS to evaluate the performance of the approaches.

We crawled these pharmaceutical websites to construct our intended datasets with the required features in a structured way. As a result, we gathered much useful information about drugs and patients' conditions and collected them into three datasets as follows:

- The first extracted dataset is the Rating dataset consists of patients' features and their ratings on drugs consisting of 3294 samples.
- The second dataset consists of Drug features containing drug categories, side effects, and benefits.
- The last dataset is the Interaction dataset containing interactions between drugs.

To evaluate the performance of our system, we used the most popular existing machine learning evaluation metrics. Accuracy, sensitivity (recall), specificity, and precision were the basic metrics that we applied to our model.

We used 70 percent of the samples (2304 samples) in the dataset for training our model, 20 percent (660 samples) for evaluation, and 10 percent (330 samples) for the test.

After obtaining the values for true positive (TP), false positive (FP), true negative (TN), and false negative (FN), different metrics can be calculated.

We compared our results with the existing approaches in [29], [50], [59], [60], and [61]. We implemented the algorithms in these papers with the datasets they have applied.

In [29], SVM and recurrent neural network (RNN) have been used to recommend a drug to a patient. In [50] the authors first considered the clustering of drugs according to the drug information, like the algorithm proposed in this paper. Then collaborative filtering is used to recommend a drug. But unlike our work, these haven't considered the classification of users and their features. Finally, in [59], an improved matrix factorization has been used, which filters the results using NSGA-III to improve the accuracy, diversity, novelty, and recall.

Table 12 represents the comparison results between our work and other drug recommendation systems in terms of important machine learning metrics.

**TABLE 12: COMPARISON RESULT OF OUR PROPOSED RECOMMENDATION SYSTEM WITH OTHER STATE-OF-THE-ART APPROACHES**

| | Accuracy | Sensitivity | Specificity | Precision | F1-Measure |
|---|---|---|---|---|---|
| SVM[26] | 0.34 | 0.75 | 0.33 | 0.04 | 0.07 |
| Neural Network [26] | 0.31 | 0.13 | 0.86 | 0.31 | 0.18 |
| Kmeans User CF [48] | 0.55 | 0.61 | 0.54 | 0.32 | 0.41 |
| NSGA III [57] | 0.63 | 0.39 | 0.66 | 0.41 | 0.39 |
| Conventional MF [58] | 0.48 | 0.45 | 0.49 | 0.33 | 0.38 |
| MLP [59] | 0.45 | 0.60 | 0.38 | 0.36 | 0.45 |
| Proposed Method | 0.65 | 0.69 | 0.64 | 0.62 | 0.65 |

Comparison results consist of the F2 measure, ROC, and confusion matrix of different approaches depicted in Figures 7 to 10.

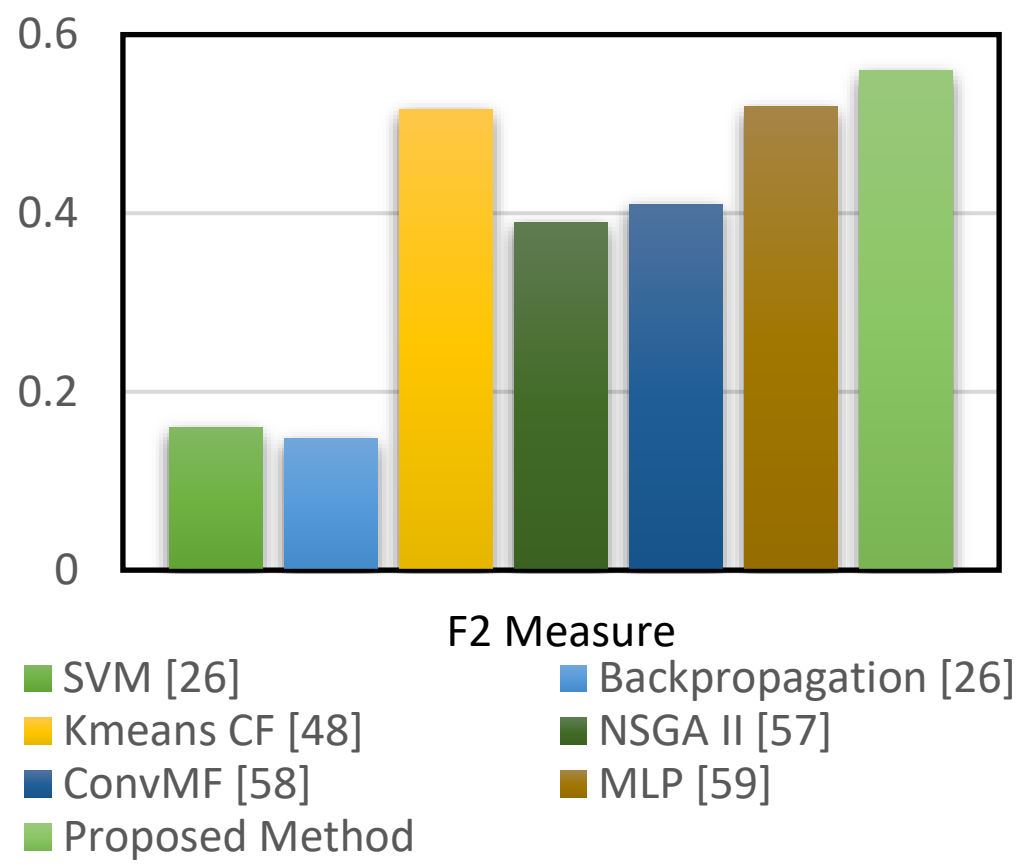

**Figure 7- Comparison result of F2 measure metric**

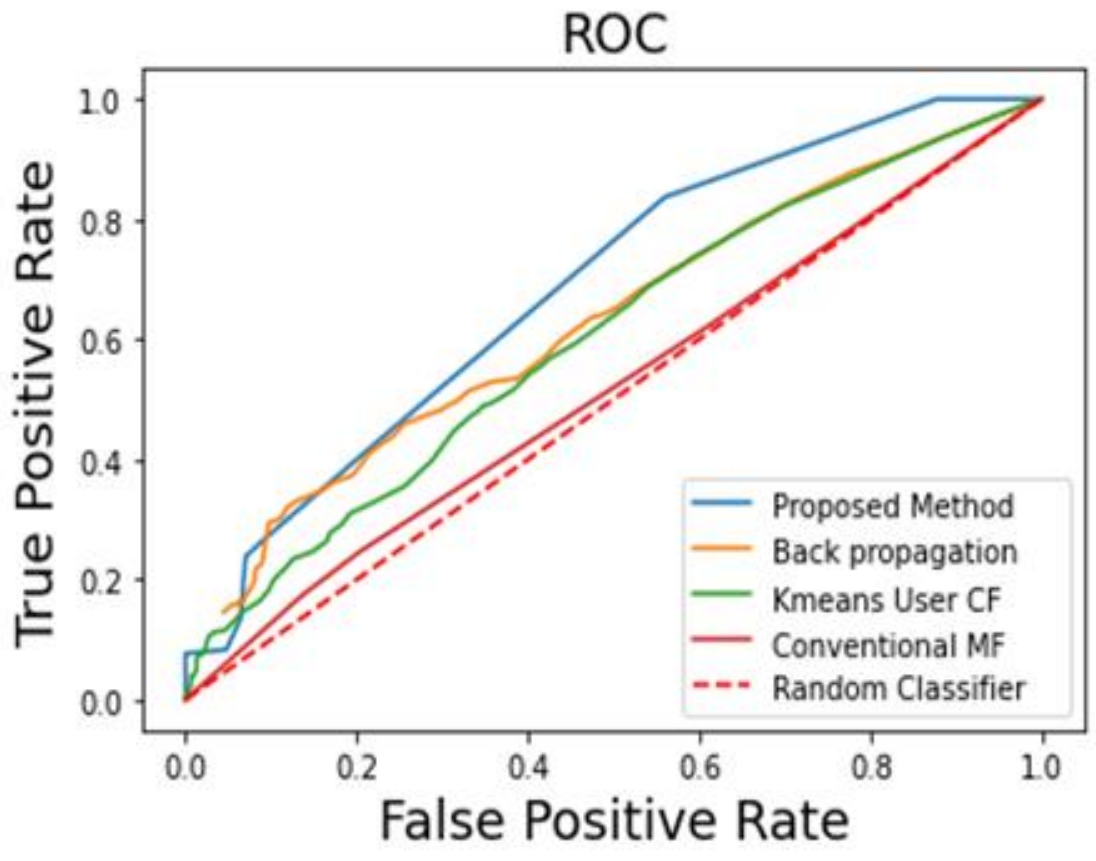

**Figure 8- comparison result of ROC.**

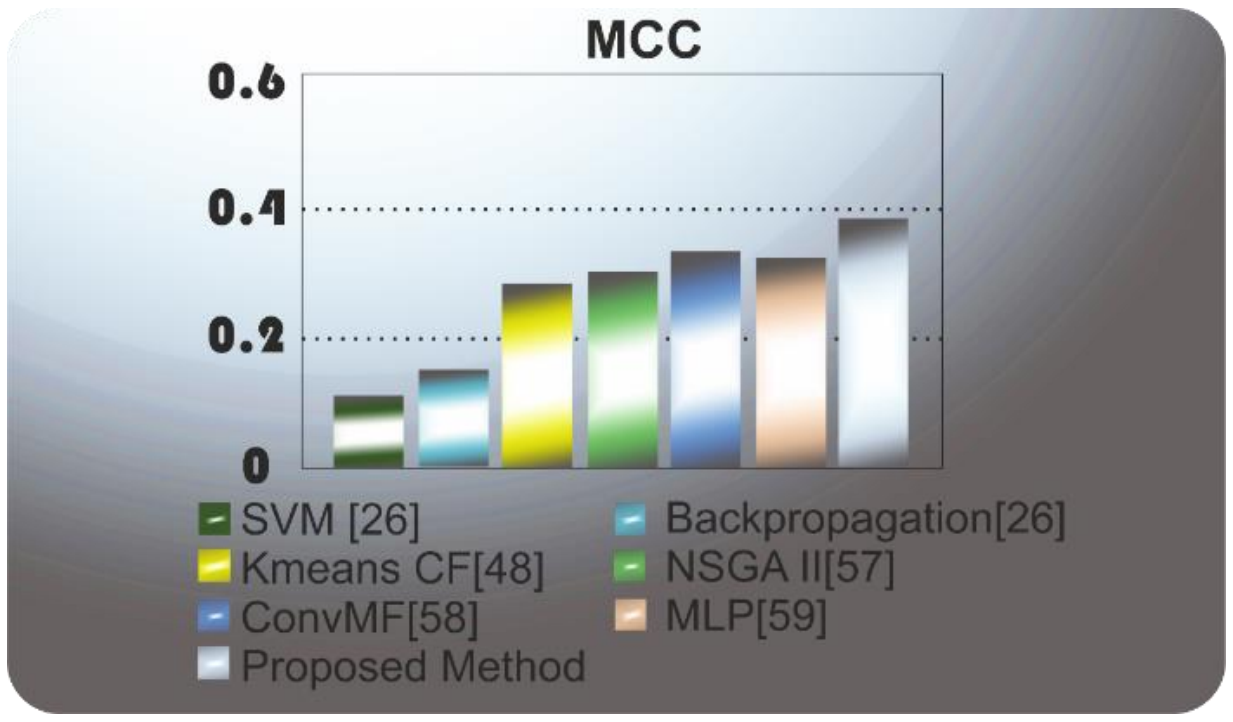

Figure 9- comparison result of MCC.

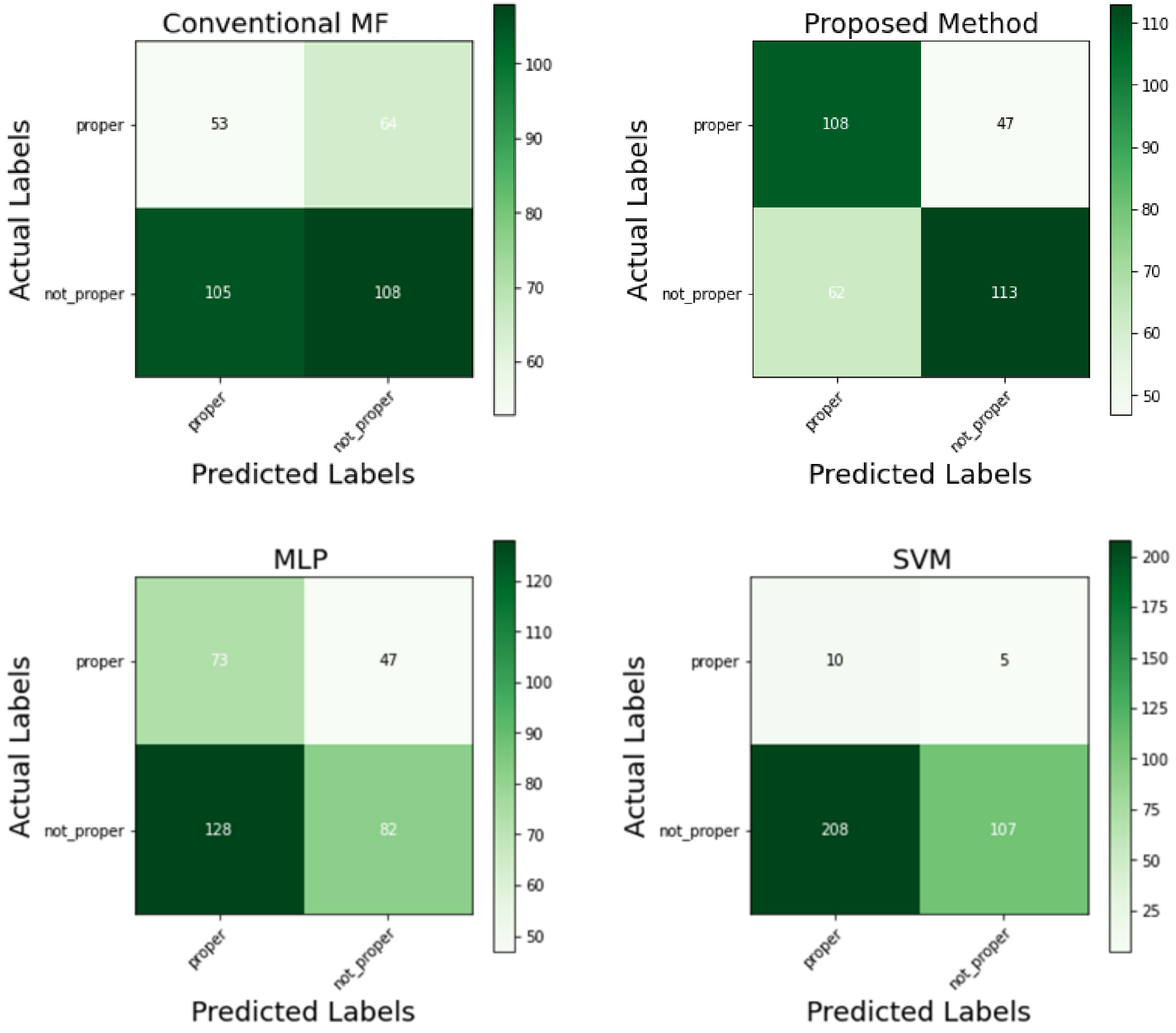

**Figure 10- Comparison result of the confusion matrix.**

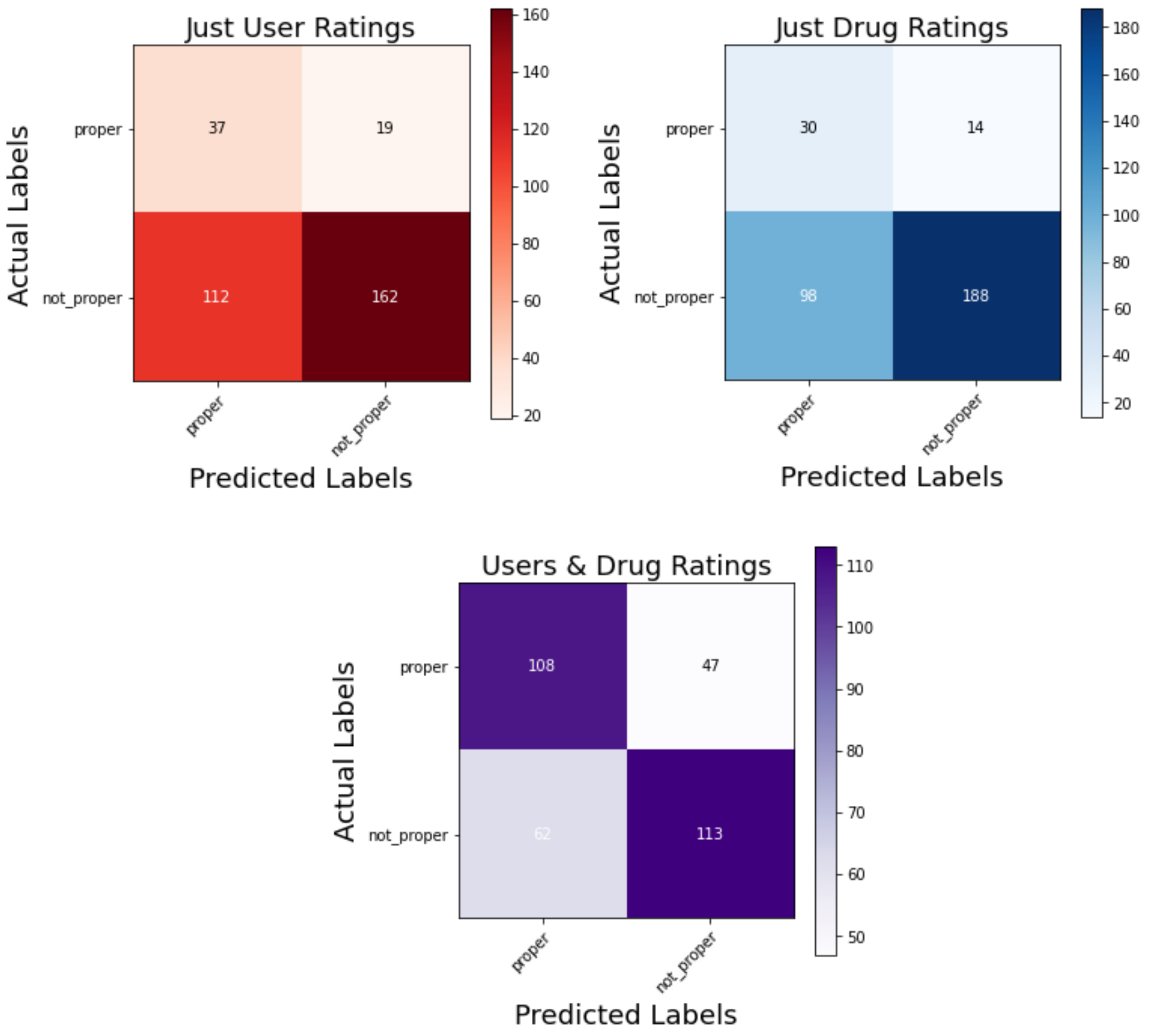

**Figure 11 -Confusion matrix obtained for the proposed method.**

The confusion matrix of predictions of our approach for user ratings compared to other approaches is shown in Figure 10. Also, the construction of confusion matrix for different ratings is also shown in Figure 11. The predictions are compared with actual ratings of users, and drugs for the case when they are considered separately and then the ratings are combined according to our proposed approach.

One of the important components of our recommender system is the final knowledge-based approach. This component prevents death, hospitalization, and disability by considering drug interactions and the user’s age. The adverse Events Dataset is used in this regard to our system's performance for recognizing such cases and recommending the appropriate drugs. This dataset contains 2486 samples, where 80% of them are used for rule extraction, and the remaining 20% are for the test.

The following parameters are considered for the evaluation:

$$Death\ Ratio = \frac{Number\ of\ Death}{Total\ Number\ of\ Recommendation}$$

$$Disability\ Ratio = \frac{Number\ of\ Disability}{Total\ Number\ of\ Recommendation}$$

$$Hospitalization\ Ratio = \frac{Number\ of\ Hospitalization}{Total\ Number\ of\ Recommendation}$$

For comparison, the system's performance for different adverse events was calculated one time without a knowledge-based component and the second time using this component.

Table 13 and Figure 12 represent the results of this comparison.

**TABLE 13- COMPARISON RESULTS OF KNOWLEDGE-BASED COMPONENT**

| Adverse event | Without knowledge-based component | With knowledge-based component |
|---|---|---|
| Death rate | 44% | 6% |
| Hospitalization | 15% | 2% |
| Disability | 4% | 0.7% |

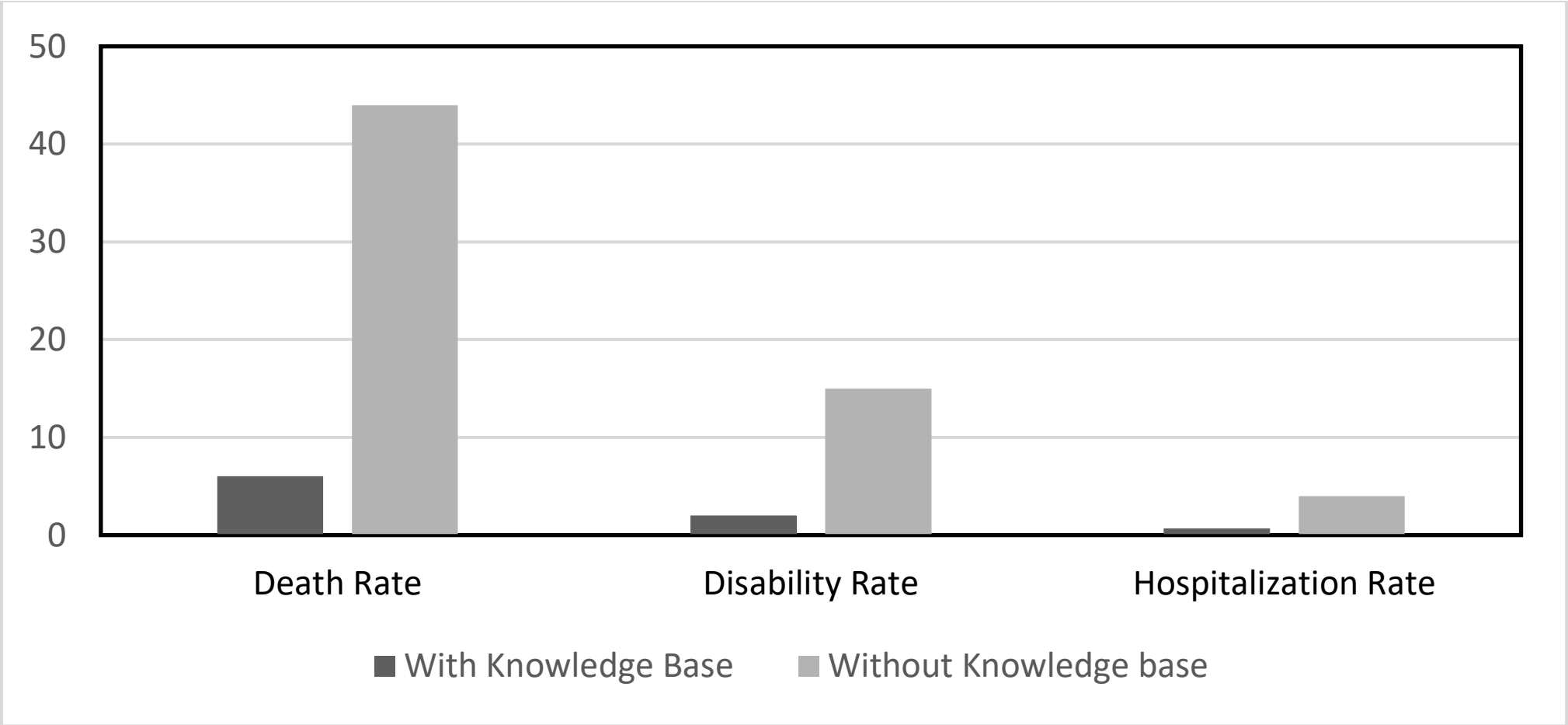

**Figure 12- Comparison result of adding knowledge-based in the recommendation system.**

Knowledge-based component is an essential part of a drug recommendation system in reducing adverse events and improving the quality of recommendations.We also considered one more important metric for recommender systems evaluations: *hit rate*.

The data set's testing samples (330) are utilized in hit-rate evaluation. The hit-rate in evaluation is calculated by the ratio of the total hits in the top 10 recommended drugs returned for all users and the total testing samples. So if $\eta$ is the number of relevant predicted drugs for all users, and $N$ is the total number of testing samples, according to [62], the hit-rate is calculated as equation (25):

$$hit_rate = \frac{\eta}{N} \quad (25)$$

The result of hit rate evaluation is represented in Figure 13. As it can be seen from this figure, our proposed approach has the $hit\ rate\ =\ 0.49$, which means it recommends more accurately than all other approaches.

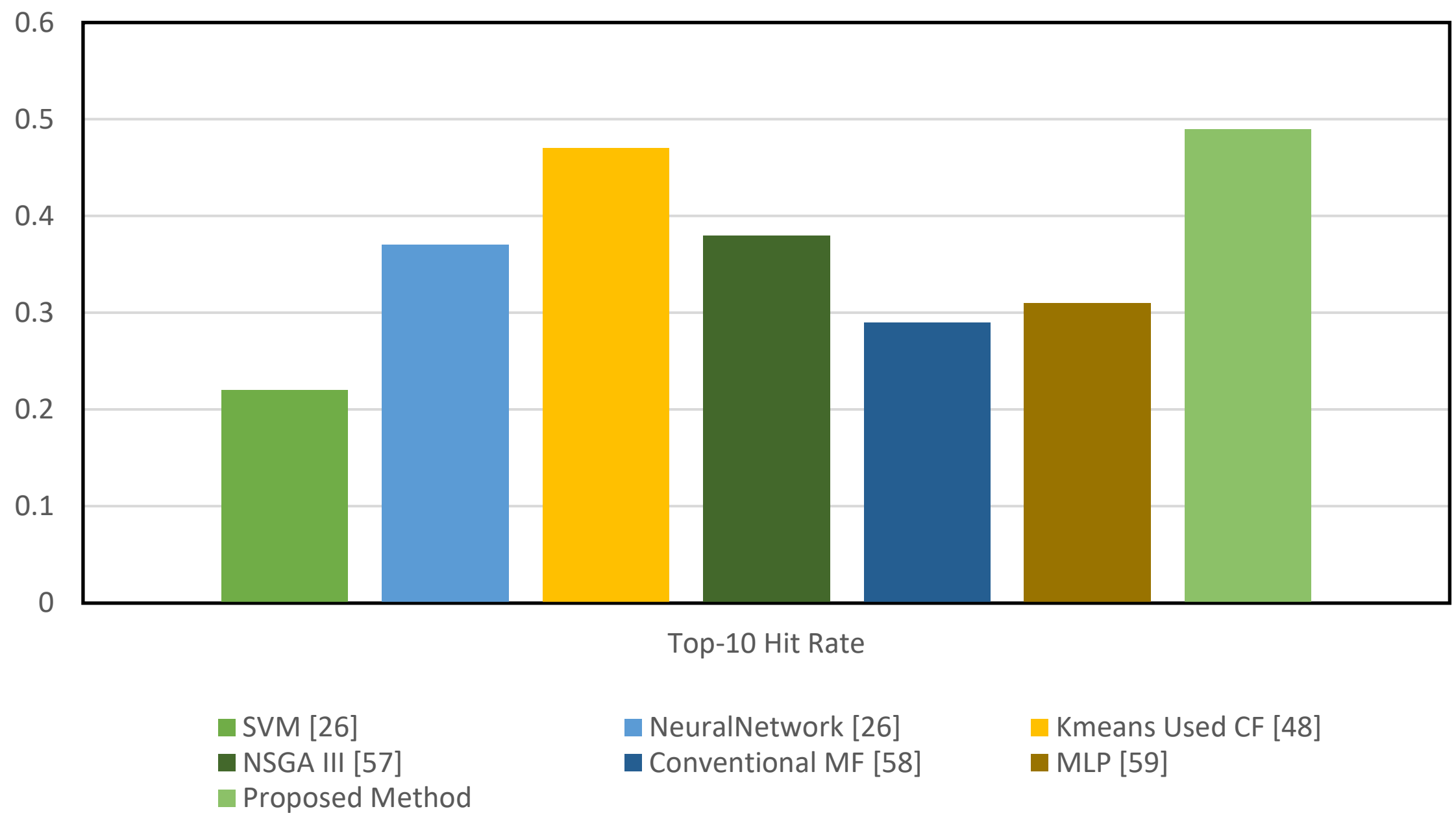

**Figure 13- Top-10 hit-rate recommendation systems.**

The next evaluation metric is *cumulative hit-rate*, which represents the number of hits with ratings above a given threshold and ignores the predicted ratings lower than the threshold. The result of the cumulative hit-rate with the threshold set to 4 is shown in Figure 14. The cumulative hit-rate is calculated as (26):

$$Cumulative\ Hit - Rate = \frac{Number\ of\ hits\ with\ rating\ above\ threshold}{N} \quad (26)$$

The utilization of this threshold makes a better match with the user's interest in the recommended drug. As our approach is a comprehensive system in terms of using different techniques and thorough datasets it is clear that we can achieve better results than other existing drug recommender systems which are limited to just one technique and using a limited dataset.

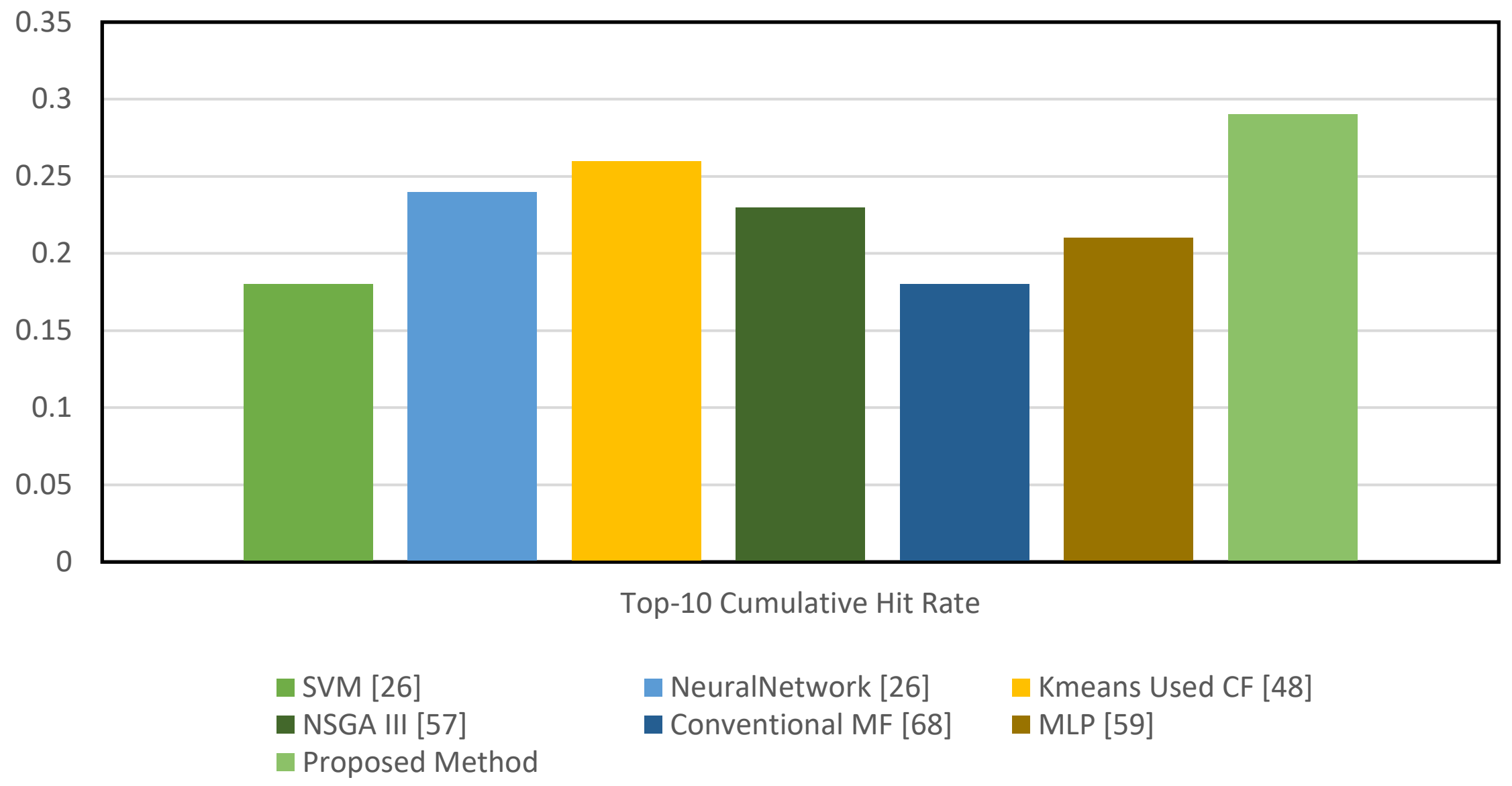

**Figure 14- Top-10 cumulative hit-rate of recommendation systems.**

## 6. DISCUSSION

Our results are encouraging in the field of drug recommendations. The model has combined the benefits of basic recommender approaches with low computational overhead through a novel modelling approach and using statistical methods. It also classifies drugs and users in terms of their features, leading to high accuracy compared to state-of-the-art algorithms. In addition, using knowledge-based component at the final stage of the recommendation system has improved significantly adverse events and the quality of recommendations. However, better results can be achieved by considering other factors like the characteristics of diseases and recommending drugs based on disease features in addition to the features of patients and drugs.

## 7. CONCLUSION

In this paper, we proposed a comprehensive drug recommender system that takes advantage of all basic recommender system techniques and applies natural language processing, neural network-based matrix factorization, and, more importantly, employs knowledge-based recommendations to recommend the most accurate drugs to patients. Compared to conventional matrix factorization, our proposed method improves the accuracy, sensitivity, and hit rate by 26%, 34%, and 40%, respectively. However it is acknowledged that this is not the same as clinical accuracy or sensitivity. In comparison with other machine learning approaches, we obtained an accuracy, sensitivity, and hit rate by an average of 31%, 29%, and 28%, respectively. Our approach can be used as an adjunct tool to recommend medicines to patients at the correct dose for them, and to reduce prescribing errors.

In the future, we will extend the knowledge and information extraction from drug databases and include all existing patient features in the user features. We will also model an integration of the features of the disease in the recommendation. These features can be captured by general practitioners, may help improve the proposed drug recommender system performance and make more accurate recommendations by having more relevant features. In the final output of the RECOMED system, we also include the dosage and effectiveness of a drug in addition to the list of drugs. Also, in our future work, we will extract the information from other drug resources like the SIDER database for drug side effects.

At the end, it should be noted that a physician would still need to review the recommendations, check for clinical sense, before prescribing the recommended medicines, for safety reasons.

## ACKNOWLEDGEMENT

We would like to express our sincere gratitude to Mohammad Beheshti Roui, whose invaluable assistance in designing the graphical figures presented in this paper greatly enhanced the visual clarity and overall quality of our work.

## DATA AVAILABILITY

The datasets generated during the current study are available in the https://github.com/DatasetsLibrary/RECOMED repositories. In addition, preprocessed datasets and source code of this study are also available at https://github.com/DatasetsLibrary/RECOMEDTool.